\documentclass[aps,twocolumn,showpacs,preprintnumbers,amsmath,amssymb,superscriptaddress,floatfix,nofootinbib]{revtex4}

\usepackage{graphicx}
\usepackage{epsfig}
\usepackage{epstopdf}
\usepackage{hyperref}
\usepackage{amsmath}
\usepackage{amsfonts}
\usepackage{amssymb}
\newcommand{\Slash}[1]{\ooalign{\hfil/\hfil\crcr$#1$}}

\begin{document}

\title{\boldmath Photoproduction of the $f_2(1270)$ resonance}
\date{\today}

\author{Ju-Jun Xie} \email{xiejujun@impcas.ac.cn}
\affiliation{Institute of Modern Physics, Chinese Academy of
Sciences, Lanzhou 730000, China} \affiliation{Research Center for
Hadron and CSR Physics, Institute of Modern Physics of CAS and
Lanzhou University, Lanzhou 730000, China} \affiliation{State Key
Laboratory of Theoretical Physics, Institute of Theoretical Physics,
Chinese Academy of Sciences, Beijing 100190, China}

\author{E.~Oset} \email{oset@ific.uv.es}
\affiliation{Institute of Modern Physics, Chinese Academy of
Sciences, Lanzhou 730000, China}\affiliation{Departamento de
F\'{\i}sica Te\'orica and IFIC, Centro Mixto Universidad de
Valencia-CSIC Institutos de Investigaci\'on de Paterna, Aptdo.
22085, 46071 Valencia, Spain}

\begin{abstract}

We have performed a calculation of the  $\gamma p \to \pi^+ \pi^- p$
reaction, where the two pions have been separated in $D$-wave
producing the $f_2(1270)$ resonance. We use elements of the local
hidden gauge approach that provides the interaction of vector mesons
in which the $f_2(1270)$ resonance appears as a $\rho$-$\rho$
molecular state in $L=0$ and spin 2. The vector meson dominance,
incorporated in the local hidden gauge approach converts a photon
into a $\rho^0$ meson and the other meson connects the photon with
the proton. The picture is simple and has no free parameters, since
the parameters of the theory have been constrained in the previous
study of the vector-vector states. In a second step we introduce new
elements, not present in the local hidden gauge approach, adapting
the $\rho$ propagator to Regge phenomenology and introducing the
$\rho NN$ tensor coupling. We find that both the differential cross
section as well as the $t$ dependence of the cross section are in
good agreement with the experimental results and provide support for
the molecular picture of the $f_2(1270)$ in the first baryonic
reaction where it has been tested.

\end{abstract}

\pacs{13.60.Le, 13.75.Lb, 14.40.Cs}

\maketitle

\section{Introduction}

The $f_2(1270)$ is a resonance that plays a prominent role in many
reactions. For instance in $\pi^0 \pi^0 \to \gamma \gamma$ it shows
with a strength much bigger than the corresponding one of the
$f_0(980)$ \cite{Marsiske:1990hx,Oest:1990ki}. From the theoretical
point of view, the $f_2(1270)$ resonance has been widely accepted as
an ordinary $q \bar q$ state belonging to a $P$-wave nonet of tensor
mesons~\cite{Klempt:2007cp,Crede:2008vw}. However, this wisdom was
challenged by the work of Ref.~\cite{raquelvec} where the
interaction of vector mesons was studied and the $f_2(1270)$ emerged
as one of the dynamically generated states from the $\rho$-$\rho$
interaction. The vector-vector interaction is described by means of
the local hidden gauge
Lagrangians~\cite{hidden1,hidden2,hidden3,hidden4} and the
unitarization of this interaction leads to vector-vector scattering
amplitudes that show poles for some states of spin $J=0,1,2$ in
$L=0$. One of them is the $f_2(1270)$. The generalization to $SU(3)$
of the $\rho$-$\rho$ interaction was done following the same lines
but with coupled channels \cite{gengvec} and the $f_2(1270)$
appeared as a dynamically generated state of the $\rho$-$\rho$
interaction with parallel spins, corroborating the findings of
Ref.~\cite{raquelvec}. The reason for the large binding of the
$f_2(1270)$ is a very large interaction of the $\rho$-$\rho$ in
$J=2$, much larger than for other channels, and larger than other
interactions known in hadron physics.

The molecular nature of $f_2(1270)$ claimed in
Refs.~\cite{gengvec,raquelvec} has been successfully tested in a
large number of processes. In Ref.~\cite{yamagata} it was shown that
it leads to a very good description of the decay rate for $f_2(1270)
\to \gamma \gamma$. It was also shown in Ref.~\cite{Branz:2009cv}
that the decay rates for two photons and one photon-one vector
decays of the $f_0(1370)$, $f_2(1270)$, $f_0(1710)$, $f'_2(1525)$
and $K^*_2(1430)$ were in good agreement with available data. A
study of the $J/\psi \to \phi(\omega) f_2(1270), ~f'_2(1525)$ and
$J/\psi \to K^{*0}(892) \bar{K}^{*0}_2(1430)$ decays was also done
within this picture in Ref.~\cite{alberzou} and good results were
obtained. Another test conducted along these lines was the radiative
decay of $J/\psi$ into $f_2(1270)$, $f'_2(1525)$, $f_0(1370)$ and
$f_0(1710)$ which was done in Ref.~\cite{Geng:2009iw} and good
results were obtained compared with the available experimental
information. Similarly, predictions for $\psi (2S)$ decay into
$\omega(\phi) f_2(1270)$, $\omega(\phi) f'_2(1525)$, $K^{*0}(892)
\bar{K}^{*\,0}_2(1430)$ and radiative decay of $\Upsilon
(1S),\Upsilon (2S), \psi (2S)$ into $\gamma f_2(1270)$, $\gamma
f'_2(1525)$, $\gamma f_0(1370)$, $\gamma f_0(1710)$ were done in
Refs.~\cite{dai2s,Dai:2015cwa} with also good agreement with
available experiments. More recently it was shown in
Ref.~\cite{xiebdec} that the ratio of the decay widths of $\bar
B^0_s \to J/\psi f_2(1270)$ to $\bar B^0_s \to J/\psi f'_2(1525)$ is
compatible with the experimental values of \cite{Aaij:2014emv}. The
nature of these states as vector meson-vector meson composite states
has undergone a large number of test than any other model. Yet, all
the test have been done in the mesonic sector and not in the
baryonic sector. The recent measurement of the photoproduction of
$f_2(1270)$ in Jefferson Lab~\cite{Battaglieri:2009aa} offers us the
first opportunity to do this new test, which we conduct here.

On the theoretical side there is a study of this reaction in
Ref.~\cite{robert}, where the idea is to create the $f_2(1270)$ as
the final state interaction of two pions in $D$-wave. Hence, one
constructs a mechanism for two pion production and then lets the
pions interact in $D$-wave to produce the $f_2(1270)$ resonance. Our
picture has some similarity with this idea in the sense that we also
generate the resonance from the final state interaction of two
mesons, but these two mesons are two $\rho$ instead of two pions.
Hence, the mechanism consists on the production of two $\rho$, for
which vector meson dominance is used, and then the two $\rho$
mesons, upon interaction, produce the $f_2(1270)$. Dealing with
baryons at a relatively large energy and large momentum transfers,
we have to introduce some elements of Regge phenomenology and
consider the $\rho NN$ tensor coupling, in addition to the vector
coupling which stems from chiral dynamics. The results obtained are
in good agreement with experiment, both for the invariant mass
dependence and the $t$ dependence, offering new support for the
picture of the $f_2(1270)$ as a $\rho$-$\rho$ molecular state.

The present paper is organized as follows. In
Sec.~\ref{sec:formalism}, we discuss the formalism and the main
ingredients of the model. In Sec.~\ref{sec:results}, we present our
main results and, finally, a short summary and conclusions are given
in Sec.~\ref{sec:summary}.

\section{Formalism and ingredients} \label{sec:formalism}

\subsection{Feynman amplitudes}

In Ref.~\cite{raquelvec} the $\rho$-$\rho$ amplitudes in $L=0$ were
classified in terms of the spin of the system, which was produced by
the $\rho$ polarization. Spin-projector operators were written in
terms of the $\rho$ polarization vectors and concretely, for spin
$S=2$, the projector was given by
\begin{eqnarray}
P^{(2)} &=& [\frac{1}{2}(\epsilon^{(1)}_i \epsilon^{(2)}_j +
\epsilon^{(1)}_j \epsilon^{(2)}_i) -\frac{1}{3}\epsilon^{(1)}_l
\epsilon^{(2)}_l \delta_{ij}] \times \nonumber \\
&& [\frac{1}{2}(\epsilon^{(3)}_i \epsilon^{(4)}_j + \epsilon^{(3)}_j
\epsilon^{(4)}_i) -\frac{1}{3}\epsilon^{(3)}_m \epsilon^{(4)}_m
\delta_{ij}] ,
\end{eqnarray}
where $\epsilon^{(k)}_i$ are the three spatial components $(i)$ of
the vector polarization of the $\rho$ meson $k$, in the order of
$\rho (1) + \rho (2) \to \rho (3) + \rho (4)$. The momenta of the
external $\rho$ mesons is assumed to be small with respect to the
mass of the $\rho$, such that the time component of the
$\epsilon^{\mu}$ ($\epsilon^0$) is neglected. This is the case of
the polarization of the photons, where we only have transverse
components. It was also discussed in Ref.~\cite{angelsphotok} that
extra terms linear in the momentum of the particles were small in
this type of processes, justifying the success of the radiative
decay, in spite of the photons not having small momenta.

\begin{figure}[htbp]
\begin{center}
\includegraphics[scale=0.7]{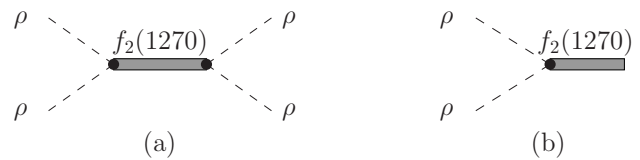}
\caption{(a): The $\rho \rho$ amplitude dominated by the $f_2(1270)$
pole; (b): Representation of the $f_2(1270)$ coupling to
$\rho\rho$.} \label{Fig:rhorhotof2}
\end{center}
\end{figure}

As we can see, the $P^{(2)}$ projector is factorized into one block
corresponding to the initial vectors and another one corresponding
to the final vectors. The amplitude close to a pole that represents
a resonance is then written as
\begin{eqnarray}
t_{\rm pole} &\simeq& \frac{g^2_T P^{(2)}_{\rm initial} P^{(2)}_{\rm
final}}{s - s_R},       \label{tpole} \\
P^{(2)}_{\rm initial} &=& \frac{1}{2}(\epsilon^{(1)}_i
\epsilon^{(2)}_j + \epsilon^{(1)}_j \epsilon^{(2)}_i)
-\frac{1}{3}\epsilon^{(1)}_l \epsilon^{(2)}_l \delta_{ij}, \\
P^{(2)}_{\rm final} &=& \frac{1}{2}(\epsilon^{(3)}_i
\epsilon^{(4)}_j + \epsilon^{(3)}_j \epsilon^{(4)}_i)
-\frac{1}{3}\epsilon^{(3)}_m \epsilon^{(4)}_m \delta_{ij},
\end{eqnarray}
where $s_R$ is the pole position and $g_T$ the coupling of the
resonance to the $\rho \rho$ component in isospin $I=0$ and spin
$S=2$. Eq.~(\ref{tpole}) is the representation of a resonance
amplitude, for instance the $f_2(1270)$ in the present case, as
shown in Fig.~\ref{Fig:rhorhotof2} (a).

Then the coupling of a resonance to $\rho \rho$ is given by the
diagram of Fig.~\ref{Fig:rhorhotof2} (b), and is expressed in terms
of the vertex~\cite{yamagata}
\begin{eqnarray}
t_{R \to \rho \rho} = g_T P^{(2)}_{\rm initial} , \label{tRrhorho}
\end{eqnarray}
with $g^2_T = 150$ GeV$^2$ that was evaluated in
Ref.~\cite{yamagata} from the $\rho \rho$ amplitude generated in
Ref.~\cite{raquelvec}.

\begin{figure}[htbp]
\begin{center}
\includegraphics[scale=1.2]{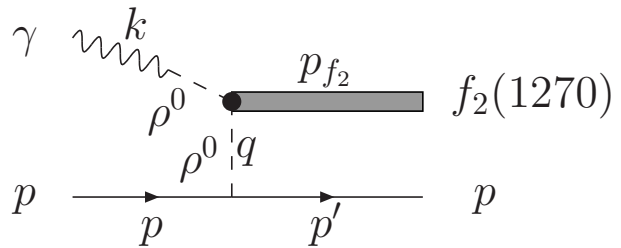}
\caption{Diagrammatic representation of the $f_2(1270)$
photoproduction. We also show the definition of the kinematical
variables ($k,~p,~p',q,p_{f_2}$) that we used in the present
calculation. In addition, we use $q=p'-p$.} \label{Fig:gammaptof2p}
\end{center}
\end{figure}

From the perspective that the $f_2(1270)$ is generated from the
$\rho \rho$ interaction, the picture for $f_2(1270)$ photoproduction
proceeds via the creation of two $\rho$ mesons by the $\gamma p$
initial state in a primary step and the interaction of the two
$\rho$ mesons generating the resonance. Alternatively we can think
that the resonance couples to two $\rho$ mesons, as depicted in
Fig.~\ref{Fig:rhorhotof2} (b) and there two $\rho$ mesons couple to
the $\gamma p$ system. This mechanism is expressed by the Feynman
diagram of Fig.~\ref{Fig:gammaptof2p}.

As we see, the photon gets converted into one $\rho^0$, a
characteristic of the local hidden gauge Lagrangian, and the
other $\rho^0$ becomes a virtual $\rho$ connecting to the
proton. Apart from the vertex of Eq.~(\ref{tRrhorho}), we need now
two ingredients, the $\gamma$-$\rho^0$ conversion vertex and the
$\rho NN$ vertex. The $\gamma$-$\rho^0$ conversion vertex is easily
obtained from the local hidden gauge
Lagrangian~\cite{hidden1,hidden2,hidden3,hidden4} (see
Ref.~\cite{Nagahiro:2008cv} for a practical set of rules) and we
have~\cite{yamagata}
\begin{eqnarray}
-i t_{\rho^0 \gamma} = \frac{-i}{\sqrt{2}} \frac{eM^2_{\rho}}{g}
\epsilon_{\mu}(\rho) \epsilon^{\mu}(\gamma), \label{trhogamma}
\end{eqnarray}
with
\begin{eqnarray}
g = \frac{M_{\rho}}{2f};~~~~f=93~~{\rm
MeV};~~~~\frac{e^2}{4\pi}=\frac{1}{137}.
\end{eqnarray}

The $\rho^0$ polarization of Eq.~(\ref{trhogamma}) is contracted
with the one of the $\rho$ in the vertex of Eq.~(\ref{tRrhorho}) and
the polarization vector of the $\rho$ in Eq.~(\ref{tRrhorho}) gets
converted into the one of the photon. The other ingredient that we
need is the vector-nucleon-nucleon vertex, which is given by the
Lagrangian
\begin{eqnarray}
{\cal L}_{BBV} = g (<\bar{B}\gamma^{\mu}[V_{\mu},B]> +
<\bar{B}\gamma^{\mu}B><V_{\mu}>), \label{lbbv}
\end{eqnarray}
which provides a $\rho^0 pp$ vertex
\begin{eqnarray}
-it_{\rho^0 pp} = i \frac{g}{\sqrt{2}} \bar{p}\gamma^{\mu} p
\epsilon_{\mu}(\rho^0).
\end{eqnarray}

On the other hand, there are generally two types of strong
couplings, vector and tensor couplings, for the $\rho NN$ vertex,
which is given in terms of the interaction Lagrangian density by
\begin{eqnarray}
{\cal L}_{\rho NN} = - g_{\rho NN} \bar{N} (\gamma^{\mu} -
\frac{\kappa_{\rho}}{2m_N}\sigma^{\mu\nu}\partial_{\nu})\vec \tau
\cdot \vec \rho_{\mu} N,
\end{eqnarray}
which gives a $\rho^0 pp$ vertex
\begin{eqnarray}
-it_{\rho^0 pp} = i g_{\rho NN} \bar{p} (\gamma^{\mu} + \frac{i
\kappa_{\rho}}{2m_N}\sigma^{\mu \nu} q_{\nu}) p
\epsilon_{\mu}(\rho^0),  \label{eq:rhonnphe}
\end{eqnarray}
with $g^2_{\rho NN}/4\pi = 0.9$ and $\kappa_{\rho} = 6.1$ obtained
from the Bonn potential~\cite{Machleidt:1987hj}. The tensor $\rho
NN$ coupling is important in describing the $\Delta^*(1620)$ and
$N^*(1535)$ production in the $\pi N$ and $NN$
collisions~\cite{Xie:2005sb,Xie:2007vs,Xie:2007qt,Ouyang:2009kv}.
Thus, we will also check its contribution in the present process.
Note that $g/\sqrt{2}$ from the local hidden gauge approach is
$2.93$, while the equivalent quantity $g_{\rho NN}$ of
Eq.~(\ref{eq:rhonnphe}) is $3.36$. They differ in $12\%$. When
including the tensor coupling we shall use the $g_{\rho NN}$
coupling of the Bonn potential~\cite{Machleidt:1987hj}.

There is one more consideration to make. The amplitude of
Eq.~(\ref{tpole}) is evaluated for a $\rho \rho$ state in the
unitary normalization, which for $I=0$ is given by (recall $|\rho^+>
= -|11>$)
\begin{eqnarray}
|\rho\rho,~I=0> = -\frac{1}{\sqrt{6}} (\rho^+ \rho^- + \rho^- \rho^+
+ \rho^0 \rho^0).
\end{eqnarray}
This has a factor $\frac{1}{\sqrt{2}}$ extra with respect to the
good normalization and is taken to  account for the identity of the
particles in the intermediate states. The good amplitude will have
$g_T$ of Eqs.~(\ref{tpole}) and (\ref{tRrhorho}) multiplied by
$\sqrt{2}$. In addition, in order to project over $\rho^0 \rho^0$ we
must multiply the coupling to the $I=0$ state by
$\frac{-1}{\sqrt{3}}$. Altogether, the coupling $\tilde{g}_T$ to be
used for $\rho^0 \rho^0$ is \footnote{In Ref.~\cite{yamagata} the
$\sqrt{2}$ factor is not implemented because it is compensated by
not dividing by two the integrated width, as it corresponds to two
final identical particles (two photons).}
\begin{eqnarray}
\tilde{g}_T = -\sqrt{\frac{2}{3}} g_T .
\end{eqnarray}

When considering photonuclear processes gauge invariance of the
amplitude is an important test, although it often happens that terms
that are relevant in the test of gauge invariance are small or
negligible in the physical
amplitudes~\cite{Borasoy:2005zg,Doring:2005bx}. Yet, in the present
case, a thorough test of gauge invariance was conducted in
Ref.~\cite{Nagahiro:2008cv} for the radiative decay of axial vector
mesons within the local hidden gauge approach, and in particular in
Ref.~\cite{yamagata} for the amplitude $\rho \rho \to \rho \gamma$,
which is the one we have here, with the two $\rho$ mesons merging
later on the $f_2(1270)$ state. The test involves terms of $\rho$
exchange in the $\rho \rho$ interaction and four body contact terms,
but all these ingredients are finally included in the effective
coupling of the $f_2(1270)$ resonance to the two $\rho$ mesons.

Considering the vertices described above, the $T$ matrix for the
diagram of Fig.~\ref{Fig:gammaptof2p} is given in the case of $\rho
NN$ vector coupling by
\begin{eqnarray}
&& -i T_{\gamma p \to f_2(1270)p} = -i \frac{e\tilde{g}_T}{2} \{
\frac{1}{2} [ \epsilon_i(\gamma) \epsilon_j(\rho) +
\epsilon_j(\gamma) \epsilon_i(\rho) ] \nonumber \\
&& -\frac{1}{3}\epsilon_m(\gamma) \epsilon_m(\rho) \delta_{ij} \}
\frac{1}{q^2 - M^2_{\rho}} <M'|\gamma^{\mu}\epsilon_{\mu}(\rho)|M>,
\label{tgammap1}
\end{eqnarray}
with $M$ and $M'$ the spin third component of the initial and final
proton. We must perform the sum over the polarizations of the $\rho$
meson exchanged in Fig.~\ref{Fig:gammaptof2p} and then we get
\begin{eqnarray}
&& T_{\gamma p \to f_2(1270)p} = \frac{e\tilde{g}_T}{2} \frac{1}{q^2
- M^2_{\rho}} [\frac{1}{2}\epsilon_i(\gamma) (-g_{j\mu} +
\frac{q_jq_{\mu}}{M^2_{\rho}}) \nonumber \\
&& + \frac{1}{2} \epsilon_j(\gamma) (-g_{i \mu} + \frac{q_i q_{\mu}}{M^2_{\rho}})  -\frac{1}{3}\epsilon_m(\gamma) \delta_{ij} (-g_{m \mu} + \frac{q_m q_{\mu}}{M^2_{\rho}})] \nonumber \\
&& <M'|\gamma^{\mu}|M>. \label{tgammap2}
\end{eqnarray}

In Eqs.~(\ref{tgammap1}) and (\ref{tgammap2}) all the components in
$g_{i\mu}$, etc, $q_i$, $q_j$, $q_{\mu}$ are covariant. The latin
indices run over $1$, $2$, $3$ and the $\mu$ index from $0, 1, 2,
3$.

In order to calculate $\bar{\sum}\sum|T|^2$ when we include the
$\rho NN$ tensor coupling, we calculate
\begin{eqnarray}
&&
\frac{1}{2}\sum_{M',M}<M'|\Gamma^{\mu}|M><M|(\Gamma^{\mu'})^\dag|M'>
\nonumber \\
&& = \frac{1}{2}
\sum_{r,r'}\bar{u}_{r'}(p')\Gamma^{\mu}u_r(p)\bar{u}_r(p)\Gamma^{\mu'}u_{r'}(p')
\nonumber \\
&& = \frac{1}{2} Tr[\frac{\Slash{p'} +
m_p}{2m_p}\Gamma^{\mu}\frac{\Slash{p} + m_p}{2m_p}\Gamma^{\mu'}]
\nonumber \\
&& = \frac{1}{8m^2_p} Tr[(\Slash{p'} + m_p)\Gamma^{\mu}(\Slash{p} +
m_p)\Gamma^{\mu'}],
\end{eqnarray}
where $\Gamma^{\mu} = \gamma^{\mu}$ or $\gamma^{\mu} + \frac{i
\kappa_{\rho}}{2m_N}\sigma^{\mu \nu} q_{\nu}$ for $\rho NN$ only
vector coupling or full vector and tensor couplings, respectively.
Hence, we have
\begin{eqnarray}
&& \bar{\sum}\sum|T|^2 = \frac{e^2\tilde{g}^2_T}{32 m^2_p (q^2 -
M^2_{\rho})^2}
\sum_{\gamma~{\rm pol.}} \sum_{i,j,m,l}\sum_{\mu,\mu'} \nonumber \\
&& [\frac{1}{2}\epsilon_i(\gamma) (-g_{j\mu} +
\frac{q_jq_{\mu}}{M^2_{\rho}})  + \frac{1}{2} \epsilon_j(\gamma)
(-g_{i \mu} + \frac{q_i q_{\mu}}{M^2_{\rho}})
\nonumber \\
&& -\frac{1}{3}\epsilon_m(\gamma) \delta_{ij} (-g_{m \mu} +
\frac{q_m q_{\mu}}{M^2_{\rho}})] [\frac{1}{2}\epsilon_i(\gamma)
(-g_{j\mu'} + \frac{q_jq_{\mu'}}{M^2_{\rho}})  \nonumber \\
&&  + \frac{1}{2} \epsilon_j(\gamma) (-g_{i \mu'} + \frac{q_i
q_{\mu'}}{M^2_{\rho}})  -\frac{1}{3}\epsilon_l(\gamma) \delta_{ij}
(-g_{l \mu'} + \frac{q_l q_{\mu'}}{M^2_{\rho}})]\nonumber \\
&& Tr[(\Slash{p'} + m_p)\Gamma^{\mu}(\Slash{p} + m_p)\Gamma^{\mu'}],
\end{eqnarray}
and we sum explicitly over all the indices and the two photon
polarization which we write explicitly as
\begin{eqnarray}
\epsilon^{(1)}(\gamma) = \left (
\begin{array}{c}
1   \\
0   \\
0
\end{array}
\right ); ~~~~\epsilon^{(2)}(\gamma) = \left ( \begin{array}{c}
0   \\
1   \\
0
\end{array} \right ),
\end{eqnarray}
where we have assumed that the photon travels in the $Z$ direction.

\subsection{Regge contributions}

In this section we explain how the Regge contributions are
implemented. We base our model on the exchange of a dominant
$\rho$-Regge trajectory, as suggested in
Refs.~\cite{Barnes:1976ek,Navelet:1976ht,Guidal:1997hy,Laget:2005be,Kochelev:2009xz}.
The $\rho$ trajectory represents the exchange of a family of
particles with $\rho$-type internal quantum numbers. In order to
take the Regge contribution into account, we replace the normal
$\rho$ meson Feynman propagator by a so-called Regge propagator,
while keeping the rest of the vertex structure, i.e.~\footnote{In
Refs.~\cite{Barnes:1976ek,Navelet:1976ht,Guidal:1997hy,Laget:2005be,Kochelev:2009xz},
trajectories with a rotating ($e^{-i \pi \alpha_{\rho}(t)}$) phase,
instead of a constant phase were assumed. This phase factor only
affects the interference between different Regge contributions,
which is not the case of the present calculation. Besides, another
phase factor, $(-1+e^{-i \pi \alpha_{\rho}(t)})/2$, was used in
Ref.~\cite{Ji:1997fb}. This later phase factor will give a dip
around $t \sim -0.7$ GeV$^2$ (corresponding to $\alpha_{\rho}(t) =
0$). Nevertheless, we will see that the CLAS data favor a constant
phase as used in Eq.~(\ref{reggepropagator}).}
\begin{eqnarray}
&& \frac{1}{q^2 - m^2_{\rho}}  ~~~~~~~~~~~~~~~({\rm normal}) \nonumber \\
& \to & \widehat{f}
(\frac{s}{s_0})^{\alpha_{\rho}(t)-1}\Gamma(1-\alpha_{\rho}(t))
~~~~({\rm Regge}),
\label{reggepropagator} \\
&& \alpha_{\rho}(t) = 0.55 + 0.8 t,
\end{eqnarray}
with $s_0 = 1$ GeV$^2$ and $\widehat{f}$ a overall normalization
factor of the Reggeon exchange contribution. This undetermined scale
will be fitted to the available data.

Note that the Regge propagator of Eq.~(\ref{reggepropagator}) has
the property that it reduces to the Feynman propagator
$1/(q^2-m^2_{\rho})$ if one approaches the first pole on the
trajectory (i.e., $q^2 \to m^2_{\rho}$). This means that the farther
we go from the pole, the more the result of the Regge model will
differ from conventional Feynman-diagram-based models.

\subsection{Differential cross section}

If we consider the $f_2(1270)$ as an elementary particle, we find
\begin{eqnarray}
\frac{d\sigma}{d\Omega} = \frac{m^2_p}{16\pi^2s}
\frac{|\vec{p}_{f_2}|}{|\vec{k}|} \bar{\sum}\sum |T|^2,
\end{eqnarray}
where $\vec{p}_{f_2}$ and $\vec{k}$ are the three momenta of final
$f_2(1270)$ and initial photon in the center of mass frame (c.m.),
and taking into account that $t= q^2 = (p-p')^2 = 2m^2_p -2
E(p)E(p') + 2 \vec{p}\cdot \vec{p'}$, we get
\begin{eqnarray}
\frac{d\sigma}{dt} = \frac{m^2_p}{16\pi s |\vec{k}|^2}
\bar{\sum}\sum |T|^2 . \label{dsigmadt}
\end{eqnarray}

\begin{figure}[htbp]
\begin{center}
\includegraphics[scale=1.2]{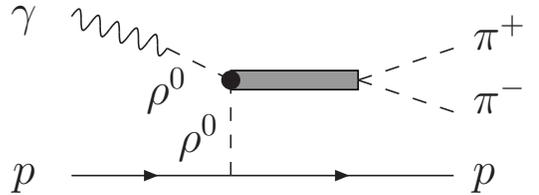}
\caption{Feynman diagram for the $\gamma p \to p \pi^+ \pi^-$
reaction.} \label{Fig:gammaptoppipi}
\end{center}
\end{figure}

The Eq.~(\ref{dsigmadt}) can be generalized for the case when the
$f_2(1270)$ is explicitly allowed to decay into two pions as shown
in Fig.~\ref{Fig:gammaptoppipi} by working out the three body phase
space and we find
\begin{eqnarray}
\frac{d^2\sigma}{dM_{\rm inv} dt} &=& \frac{m^2_p}{8\pi^2 s
|\vec{k}|^2} \frac{M^2_{\rm inv} \Gamma_{\pi}}{|M^2_{\rm inv} -
M^2_{f_2} + i
M_{\rm inv} \Gamma_{f_2}|^2} \nonumber \\
&& \times \bar{\sum}\sum |T|^2, \label{dsigmadmdt}
\end{eqnarray}
where $M_{\rm inv}$ is the invariant mass distribution of the two
pions, $\Gamma_{f_2}$ is the total decay width of the $f_2(1270)$
and $\Gamma_{\pi}$ is the partial decay width of the $f_2(1270)$
into the $\pi \pi$ system, in our case $\pi^+\pi^-$. The $\pi^+
\pi^-$ decay accounts for $\frac{2}{3}$ of the $\pi\pi$ decay width
of the $f_2(1270)$ which is $85\%$ of $\Gamma_{f_2}$. Since the
$\pi\pi$ decay is in $D$-wave, in order to have $\Gamma_{\pi}$ and
$\Gamma_{f_2}$ in the range of invariant masses that we consider
(close to the $f_2(1270)$ resonance), we take
\begin{eqnarray}
&& \Gamma_{\pi}(M_{\rm inv}) = \Gamma_{\pi}^{\rm on} (
\frac{\tilde{q}}{\bar{q}} )^5 \frac{M^2_{f_2}}{M^2_{\rm inv}} , \\
&& \Gamma_{f_2} (M_{\rm inv}) = 0.85 \Gamma_{f_2}^{\rm on} (
\frac{\tilde{q}}{\bar{q}} )^5 \frac{M^2_{f_2}}{M^2_{\rm inv}} + 0.15
\Gamma_{f_2}^{\rm on},
\end{eqnarray}
with $\Gamma^{\rm on}_{f_2} = 185$ MeV, $\Gamma^{\rm on}_{\pi} =
105$ MeV, and $M_{f_2} = 1275$ MeV~\cite{Agashe:2014kda}. And
\begin{eqnarray}
\tilde{q} &=& \frac{\lambda^{1/2}(M^2_{\rm inv}, m^2_{\pi},
m^2_{\pi})}{2M_{\rm inv}}, \\
\bar{q} &=& \frac{\lambda^{1/2}(M^2_{f_2}, m^2_{\pi},
m^2_{\pi})}{2M_{f_2}},
\end{eqnarray}
where $\lambda$ is the K\"allen function with $\lambda(x,y,z) =
(x-y-z)^2 - 4yz$.

This kind of parametrization of the width of a resonance is common
and it is meant to take into account the phase space of each decay
mode as a function of the
energy~\cite{gengvec,Chiang:1990ft,Hanhart:2010wh}. In the present
case we take explicitly the phase space for the $D$-wave decay of
the $f_2(1270)$ into two pions, and for the rest of the width, which
accounts for only $15\%$ of the width, coming mostly from four pion
decay, we have made an approximation and maintained it constant.

\subsection{Background}

As in Ref.~\cite{Battaglieri:2009aa}, we consider an incoherent
background contribution to the di-pion mass spectrum of the $\gamma
p \to p \pi^+ \pi^-$ reaction as~\footnote{In general, the di-pion
mass distribution of $\gamma p \to p \pi^+ \pi^-$ reaction can be
obtained by working out the three body phase space,
$\frac{d^2\sigma}{dM_{\rm inv}dt} = \frac{m^2_p}{2^8\pi^3 s
|\vec{k}|^2} \sum |{\cal M}|^2 \tilde{q}$, with ${\cal M}$ the
scattering amplitude of the $\gamma p \to p \pi^+ \pi^-$ reaction.}
\begin{eqnarray}
\frac{d^2\sigma}{dM_{\rm inv}dt} = \frac{C m^2_p}{2^8\pi^3 s
|\vec{k}|^2} \tilde{q},
\end{eqnarray}
where $C$ is constant.

\section{Numerical results} \label{sec:results}

Here we present our numerical results. First, we label three models
in Table~\ref{Tab:models}. Model A represents our calculation
including only the vector $\rho NN$ coupling with the normal $\rho$
meson propagator. Model B includes the vector and tensor $\rho NN$
couplings with the normal $\rho$ meson propagator. Model C
represents the calculation including the vector and tensor $\rho NN$
couplings with the $\rho$-Regge propagator.

\begin{table}[htbp]
\caption{Relevant combinations in Models A, B and C.}
\begin{center}
\begin{tabular}{ccc}
\hline\hline
& $\rho NN$ vertex & $\rho$ propagator  \\
\hline
\rm{Model A} & vector  & normal  \\
\rm{Model B} & vector + tensor & normal   \\
\rm{Model C} & vector + tensor & Regge    \\
\hline \hline
\end{tabular}
\end{center} \label{Tab:models}
\end{table}

\begin{figure}[htbp]
\begin{center}
\includegraphics[scale=0.45]{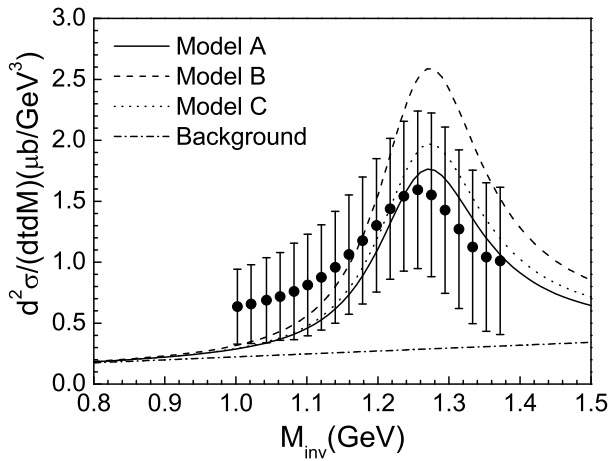}
\caption{Theoretical prediction for $D$-wave $\pi \pi$ mass
distribution at $E_{\gamma} = 3.3$ GeV and $t=-0.55$ GeV$^2$
compared with the CLAS data taken from
Ref.~\cite{Battaglieri:2009aa}. Models A, B and C are explained in
the text.} \label{Fig:dsigdmdt}
\end{center}
\end{figure}

In Ref. \cite{Battaglieri:2009aa} the $\gamma p \to \pi^+ \pi^- p$
reaction was studied in the photon energy range $3.0-3.8$ GeV. The
magnitude $\frac{d\sigma}{dM_{\rm inv} d\Omega_{\pi} dt}$ was
measured, where $M_{\rm inv}$ is the invariant mass of the two
pions, $t$ the momentum transfer squared and $\Omega_{\pi}$, the
angle of one pion measured in the $\pi^+ \pi^-$ helicity rest frame.
This multiple differential cross section was then projected over the
appropriate partial waves with the integral over the corresponding
spherical-harmonic functions. After correcting for detector
acceptance and detector efficiency, $\frac{d\sigma}{dM_{\rm inv}
dt}$ for the corresponding partial waves was produced, such that a
direct comparison with a theoretical models can be done. The
projection over $\pi^+ \pi^-$ $D$-wave was done and a neat peak
around 1270 MeV, corresponding to the $f_2(1270)$ resonance, was
found (see Fig. \ref{Fig:dsigdmdt}) for a certain cut of the photon
energy and the variable $t$.

Simultaneously, a photon energy averaged cross section
$\frac{d\sigma}{ dt}$ was determined by considering a wide range of
energies and integrating $\frac{d\sigma}{dM_{\rm inv}dt}$ over
$M_{\rm inv}$ around the resonance peak. These are magnitudes that
we can easily address with our theoretical framework and we show
below the results obtained.

In Fig.~\ref{Fig:dsigdmdt} we shown the results of Models A, B and C
for $\frac{d\sigma}{dM_{\rm inv}dt}$ as a function of $M_{\rm inv}$
for $E_{\gamma} = 3.3$ GeV and $t = -0.55$ GeV$^2$, which we compare
with the results of Ref.~\cite{Battaglieri:2009aa} measured in the
range $3.2 < E_{\gamma} < 3.4$ GeV, $-0.6 < t < -0.5$ GeV$^2$. The
result shown in Fig.~\ref{Fig:dsigdmdt} of Model C is obtained with
$\widehat{f} = 1.43$, while for the case of Model B, we need a
dipole form factor $(\frac{\Lambda^2}{\Lambda^2 - q^2})^2$ for the
$\rho NN$ vertex to suppress the contribution of high momenta. The
result shown in Fig.~\ref{Fig:dsigdmdt} of Model B is obtained with
$\Lambda = 0.75$ GeV. For Model A, all the parameters were
determined before as discussed in previous section. Besides, in
Fig.~\ref{Fig:dsigdmdt} we also show the results for the background
with the dash-dotted line. As stated in
Ref.~\cite{Battaglieri:2009aa}, a smooth background is considered in
the analysis which is introduced by the reflection of baryon
resonances and is expected to be smooth and structureless,
contributing to all partial waves. We show in
Fig.~\ref{Fig:dsigdmdt} that a smooth background that we consider of
$S$-wave for simplicity, meant to account for the lower side of
invariant masses where our model lacks some strength, would improve
the agreement with the data. We do not elaborate further since this
contribution is small compared to the signal of the resonance in the
region of the peak.

As we can see, the experimental data have a wide band of allowed
values, but our theoretical results of all three models around the
peak go through the middle of the band. Discrepancies below the
resonance peak can be attributed to background which we have not
considered in our approach, since only the resonance contribution is
taken into account.

\begin{figure}[htbp]
\begin{center}
\includegraphics[scale=0.45]{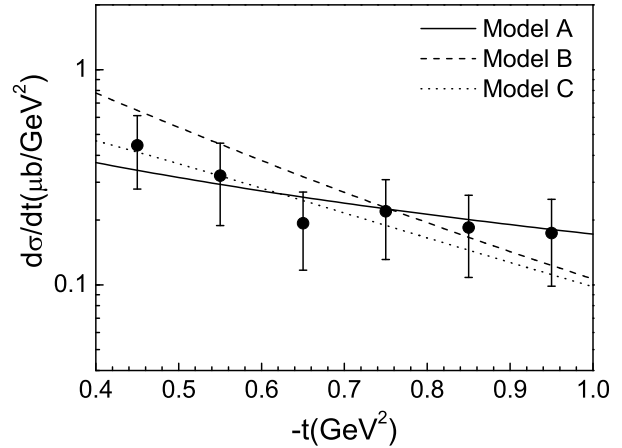}
\caption{Differential cross section $\frac{d\sigma}{dt}$ as a
function of $t$. The experimental data are taken from
Ref.~\cite{Battaglieri:2009aa}. The small background of
Fig.~\ref{Fig:dsigdmdt} is not considered.} \label{Fig:dsigdt}
\end{center}
\end{figure}

Further information can be obtained from the $t$ dependence of the
cross section. In Ref.~\cite{Battaglieri:2009aa} one finds
$\frac{d\sigma}{dt}$ as a function of $t$, obtained by integrating
$M_{\rm inv}$ in the range $1.275 \pm 0.185$ GeV and for photon
energy in the range $E_{\gamma} = 3.0 - 3.8$ GeV. We perform the
calculation for this observable and the results are shown in
Fig.~\ref{Fig:dsigdt}. Once again, all the three models give good
agreements with experiment.

Although the data seem to imply a slightly bigger slope than
provided by the Model A which only the vector $\rho NN$ coupling is
included, the fact is that the results, with no free parameters,
agree well with the data within errors. It seems clear that the
slope provided by the $\rho$ meson propagator in Model A accounts
for the bulk of the $t$ dependence of the cross section, but one
cannot exclude extra elements to the theory that gradually change
the cross section at larger values of $t$, too large to be
accommodated within an effective theory as we have used. The Model C
including both vector and tensor $\rho NN$ couplings with the
$\rho$-Regge propagator gives similar results with Model A. Yet, for
the range of energies and momentum transfers measured, it looks
clear that Model A provides the basic features of the experiment and
globally we can claim a good agreement with experiment.

\begin{figure}[htbp]
\begin{center}
\includegraphics[scale=0.45]{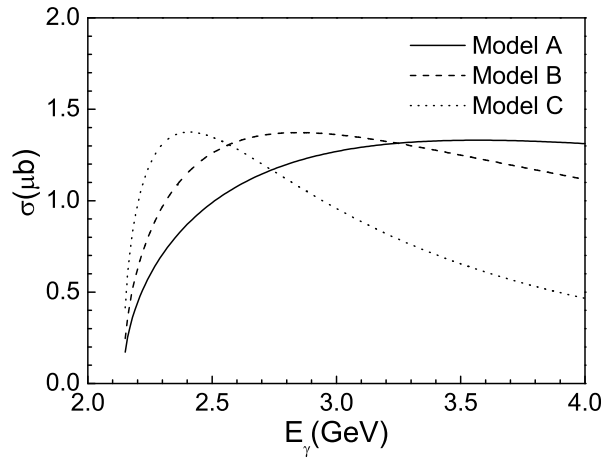}
\caption{Total cross section for the $\gamma p \to p f_2(1270)$
reaction as a function of $E_{\gamma}$. The small background of
Fig.~\ref{Fig:dsigdmdt} is not considered.} \label{Fig:tcs}
\end{center}
\end{figure}

In addition to the differential cross section, we calculate also the
total cross section for the $\gamma p \to p f_2(1270)$ reaction as a
function of the photon beam energy $E_{\gamma}$. The theoretical
results are shown in Fig.~\ref{Fig:tcs}. We see that there is a
clear bump structure around $E_{\gamma} = 2.4$ GeV for Model C,
while for Model B, this bump structure is much weaker than Model C,
but stronger than Model A. Therefore, this observable, the total
cross section of the $\gamma p \to p f_2(1270)$ reaction can be
employed, in the future experiments at CLAS, to test our model
calculation.

\section{Conclusions} \label{sec:summary}

In this paper we have studied the $\gamma p \to \pi^+ \pi^- p$
reaction performed in Jefferson Lab, where the two pions have been
separated in $D$-wave, producing the $f_2(1270)$ resonance. This
resonance has been much studied recently from the point of view of a
$\rho \rho$ molecule and has passed all tests in the reactions where
it has been studied. Yet, all the reactions were mesonic reactions.
This is the first time where this idea has been tested in a baryonic
reaction. The elements needed for the test are very simple, which
offers a special transparency in the interpretation of the results.
On the one side the $f_2(1270)$ couples to $\rho \rho$ in $I=0$ and
the value of the coupling has been obtained before in the theory
that provides the $f_2(1270)$ as a $\rho \rho$ molecule based on the
local hidden gauge formalism for the interaction of vector mesons.
With this coupling and the vector meson dominance hypothesis,
incorporated in the local hidden gauge approach, the photon gets
converted into one of the $\rho^0$ of the $\rho \rho$ formation
state of the $f_2(1270)$, and the other $\rho^0$ acts as a mediator
between the photon and the proton.

With this simple picture we determine both the differential cross
section and the $t$ dependence of the integrated cross section over
the invariant mass around the resonance with three different models,
only vector $\rho NN$ coupling is considered, both vector and tensor
$\rho NN$ couplings are considered, and the $\rho$-Regge propagator
is also checked. All these three models can reproduce the
experimental data. The agreement with the experimental differential
cross sections and the $t$ dependence is good, thus, providing new
support for the $\rho \rho$ molecular picture of the $f_2(1270)$.
Furthermore, we calculate also the total cross section of the
$\gamma p \to p f_2(1270)$ reaction, it is found that the shapes of
those three models are different, and our model calculation could be
teset by the future experiments at CLAS.

\section*{Acknowledgments}

One of us, E. O., wishes to acknowledge support from the Chinese
Academy of Science (CAS) in the Program of Visiting Professorship
for Senior International Scientists (Grant No. 2013T2J0012). This
work is partly supported by the Spanish Ministerio de Economia y
Competitividad and European FEDER funds under the contract number
FIS2011-28853-C02-01 and FIS2011-28853-C02-02, and the Generalitat
Valenciana in the program Prometeo II-2014/068. We acknowledge the
support of the European Community-Research Infrastructure
Integrating Activity Study of Strongly Interacting Matter (acronym
HadronPhysics3, Grant Agreement n. 283286) under the Seventh
Framework Programme of EU. This work is also partly supported by the
National Natural Science Foundation of China under Grant Nos.
11105126 and 11475227.

\bibliographystyle{plain}

\end{document}